\documentclass[a4paper,12pt]{article}
\usepackage{graphicx,rotating,hyperref,slashed,amsmath,xcolor,amssymb,amsfonts}
\pdfoutput=1
\makeatletter
\hypersetup{colorlinks,bookmarksopen,bookmarksnumbered,
linkcolor=blus,pdfstartview=FitH,urlcolor=rossos,citecolor=verde}
\allowdisplaybreaks

\def\lsim{\mathrel{\rlap{\lower3pt\hbox{\hskip0pt$\sim$}}
   \raise1pt\hbox{$<$}}}         
\def\gsim{\mathrel{\rlap{\lower4pt\hbox{\hskip1pt$\sim$}}
   \raise1pt\hbox{$>$}}}         

\newcommand{\mio}[1]{}

\newcommand{\fig}[1]{~\ref{fig:#1}}

\definecolor{Gray}{gray}{0.95}

\newcommand{\X}{S}

\newcommand{\fb}{\,{\rm fb}}

\usepackage{multicol}
\usepackage{color}
\definecolor{rosso}{cmyk}{0,1,1,0.4}
\definecolor{rossos}{cmyk}{0,1,1,0.55}
\definecolor{rossoc}{cmyk}{0,1,1,0.2}
\definecolor{blu}{cmyk}{1,1,0,0.3}
\definecolor{blus}{cmyk}{1,1,0,0.6}
\definecolor{bluc}{cmyk}{1,1,0,0.1}
\definecolor{verde}{cmyk}{0.92,0,0.59,0.25}
\definecolor{verdec}{cmyk}{0.92,0,0.59,0.15}
\definecolor{verdes}{cmyk}{0.92,0,0.59,0.4}

\newcommand{\eq}[1]{~{\rm (\ref{eq:#1})}}

\newcommand{\GeV}{\,{\rm GeV}}
\newcommand{\TeV}{\,{\rm TeV}}

\def\circa#1{\,\raise.3ex\hbox{$#1$\kern-.75em\lower1ex\hbox{$\sim$}}\,}

\newcommand{\beq}{\begin{equation}}
\newcommand{\eeq}{\end{equation}}

\newcommand{\bea}{\begin{eqnarray}}
\newcommand{\eea}{\end{eqnarray}}
\newcommand{\be}{\begin{equation}}
\newcommand{\ee}{\end{equation}}
\font\tenrsfs=rsfs10 at 12pt
\font\sevenrsfs=rsfs7 at 10 pt
\font\fiversfs=rsfs5
\newfam\rsfsfam
\textfont\rsfsfam=\tenrsfs
\scriptfont\rsfsfam=\sevenrsfs
\scriptscriptfont\rsfsfam=\fiversfs
\def\mathscr#1{{\fam\rsfsfam\relax#1}}

\def\Lag{\mathscr{L}}

\def\circa#1{\,\raise.3ex\hbox{$#1$\kern-.75em\lower1ex\hbox{$\sim$}}\,}
\makeatletter

\def\hhref#1{\href{http://arxiv.org/abs/#1}{arXiv:#1}} 

\def\hhref#1{\href{http://arxiv.org/abs/#1}{arXiv:#1}} 
 
\def\art{\@ifnextchar[{\eart}{\oart}}
\def\eart[#1]#2#3#4#5#6{{\rm #2}, {\em #3 \bf #4} {\rm (#6) #5} ({\em #1})}

\def\article{\@ifnextchar[{\earticle}{\oarticle}}
\def\oarticle#1#2#3#4#5#6{{\rm #1}, {\em ``#6''}, {\rm #2 #3 (#5) #4}}
\def\earticle[#1]#2#3#4#5#6#7{{\rm #2}, {\em ``#7''}, {\rm #3 #4 (#6) #5}  [\hhref{#1}]}
\def\hepart[#1]#2{{\rm #2, \em#1}}
\def\heparticle[#1]#2#3{#2, {\em ``#3''} [\hhref{#1}]}

\newcounter{alphaequation}[equation]
\def\thealphaequation{\theequation\hbox to
0.6em{\hfil\alph{alphaequation}\hfil}}
\def\eqnsystem#1{
\def\@eqnnum{{\rm (\thealphaequation)}}
\def\@@eqncr{\let\@tempa\relax \ifcase\@eqcnt \def\@tempa{& & &} \or
  \def\@tempa{& &}\or \def\@tempa{&}\fi\@tempa
  \if@eqnsw\@eqnnum\refstepcounter{alphaequation}\fi
\global\@eqnswtrue\global\@eqcnt=0\cr}
\refstepcounter{equation} \let\@currentlabel\theequation \def\@tempb{#1}
\ifx\@tempb\empty\else\label{#1}\fi
\refstepcounter{alphaequation}
\let\@currentlabel\thealphaequation
\global\@eqnswtrue\global\@eqcnt=0 \tabskip\@centering\let\\=\@eqncr
$$\halign to \displaywidth\bgroup \@eqnsel\hskip\@centering
$\displaystyle\tabskip\z@{##}$&\global\@eqcnt\@ne
\hskip2\arraycolsep\hfil${##}$\hfil& \global\@eqcnt\tw@\hskip2\arraycolsep
$\displaystyle\tabskip\z@{##}$\hfil
\tabskip\@centering&\llap{##}\tabskip\z@\cr}
\def\endeqnsystem{\@@eqncr\egroup$$\global\@ignoretrue} \makeatother

\newcommand{\SU}{\,{\rm SU}}

\newcommand{\Ggg}{\Gamma_{\gamma\gamma}}
\definecolor{fiorentina}{rgb}{.5,0,.5}

\begin{document}

\centerline{CERN-TH-2016-028  \hfill IFUP-TH/2016}
\bigskip
\bigskip

\begin{center}
{\LARGE \bf \color{rossos} On the maximal diphoton width}\\[1cm]

\bigskip\bigskip

{\large\bf Alberto Salvio$^a$, Florian Staub$^a$,}\\[3mm]
{\large\bf Alessandro Strumia$^{a,b}$, Alfredo Urbano$^{a}$
}
\\[5mm]

\bigskip

{\it $^a$ Theoretical Physics Department, CERN, Geneva, Switzerland}\\[1mm]
{\it $^b$ Dipartimento di Fisica dell'Universit{\`a} di Pisa and INFN, Italy}

\bigskip

\vspace{1cm}
{\large\bf\color{blus} Abstract}
\begin{quote}\large
Motivated by the 750 GeV diphoton excess found at LHC,
we compute the maximal width into $\gamma\gamma$  that a neutral scalar 
can acquire through a loop of charged fermions or scalars
as function of the maximal scale at which the theory holds,
taking into account vacuum (meta)stability bounds.
We show how an extra gauge symmetry can qualitatively weaken such bounds,
and explore collider probes and connections with Dark Matter.
\end{quote}
\thispagestyle{empty}
\end{center}

\setcounter{page}{1}
\setcounter{footnote}{0}

\tableofcontents

\newpage

\section{Introduction}
The ATLAS and CMS collaborations found an excess in
 $pp\to \gamma\gamma$ events~\cite{data} that can be interpreted as the production of a new scalar resonance $S$ with mass
$M_S\approx 750\GeV$, provided that $S$ has a large enough width into photons, $\Gamma_{\gamma\gamma}=\Gamma(S\to\gamma\gamma)$.
Assuming that $S$ is produced trough $gg$ or $q\bar q$ partonic collisions,
the claimed $\gamma\gamma$ excess can be reproduced for $\Gamma_{\gamma\gamma}/M_S\approx 10^{-6}$ if the $S$ width is narrow,
and for $\Gamma_{\gamma\gamma}/M\approx 10^{-4}$ if the total width is large, $\Gamma_S\sim 0.06 M_S$.
Larger values $\Gamma_{\gamma\gamma}/M\approx 10^{-3}$ are needed if $S$ is produced trough
$\gamma\gamma$ partonic collisions~\cite{750fits,750others}.

\medskip

This raises a theoretical question: how can  such a width be obtained in a fundamental theory?
Extra charged fermions or scalars $X$ must be present to mediate the $S\to\gamma\gamma$ process,
and they must be coupled to $S$, through Yukawa couplings  $y$ or scalar cubic couplings $\kappa$.

\medskip

In the fermionic case,
$\Ggg$ gets enhanced by considering a large Yukawa $y$ and/or a large multiplicity $N$ and/or a large hypercharge $Y$
of the new fermions.
All these enhancements imply that some coupling ($y$ and/or $g_Y$), when renormalised up to higher energies, 
becomes larger until it develops a Landau pole,
signalling the presence of new non-perturbative physics~\cite{Kam,Urb,750fits}.
In section~\ref{psi} we revisit such issues, adding the extra constraint of vacuum stability along the $S$ direction.

\medskip

In the scalar case, the loop that mediates $S\to \gamma\gamma$
can be enhanced by a large cubic $\kappa S|X|^2$~\cite{Raidal,trini}.
At first sight, this presents two possible advantages.
First, the RGE evolution of $\kappa$ never generates Landau poles
since it has dimension 1 and thereby corresponds to a relevant operator
(unlike the dimensionless Yukawa coupling $y$ introduced in the fermionic case).
Furthermore, a large cubic can arise if there is a weakly-coupled scalar sector 
around $\approx10 \TeV$ that contains the accidentally light scalars $S$ and $X$
with a cubic coupling among them which does not get accidentally suppressed.
However, a large cubic leads to extra minima in the potential $V(S,X)$ and is thereby subject to vacuum stability bounds.
In this work we consider absolute stability 
and meta-stability.
We will find that, after imposing such bounds, the maximal $\Ggg$ given by a scalar loop is similar to
the maximal $\Ggg$ produced by a fermion loop.

In section~\ref{psi} we reconsider fermion models.
In section~\ref{min} we consider  scalar models.
Signals at colliders and connection with Dark Matter is discussed in section~\ref{col}.
Conclusions are given in section~\ref{concl}.

\begin{figure}[t]
\begin{center}
$$\includegraphics[width=0.45\textwidth]{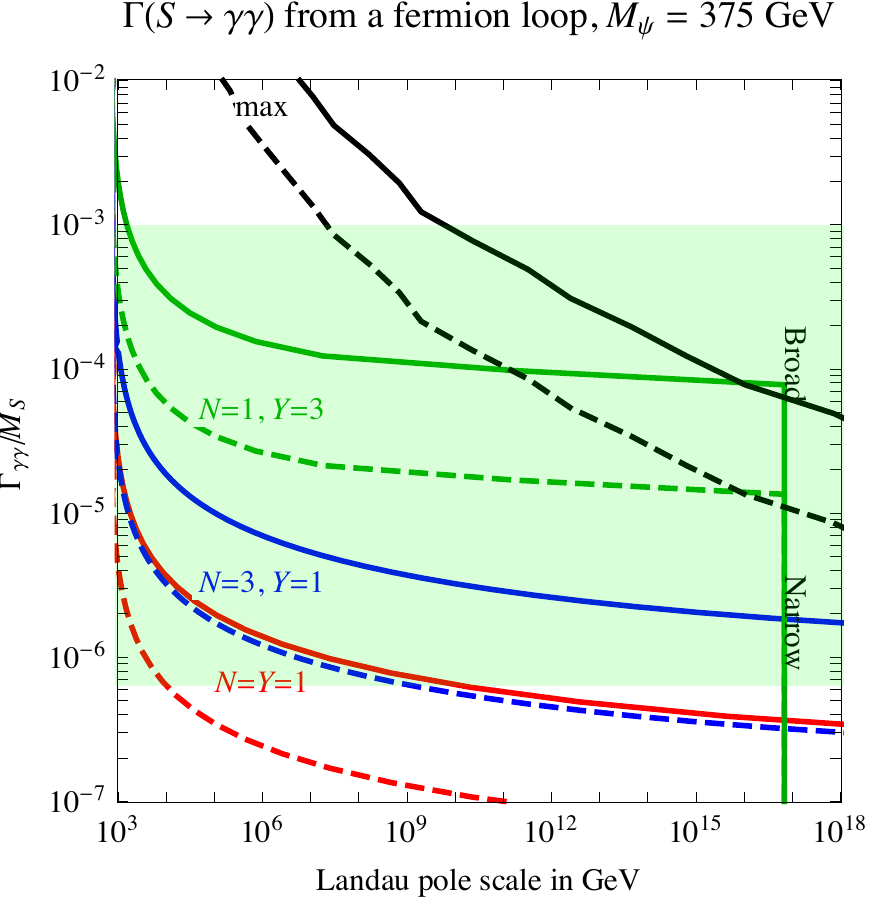}\qquad\includegraphics[width=0.45\textwidth]{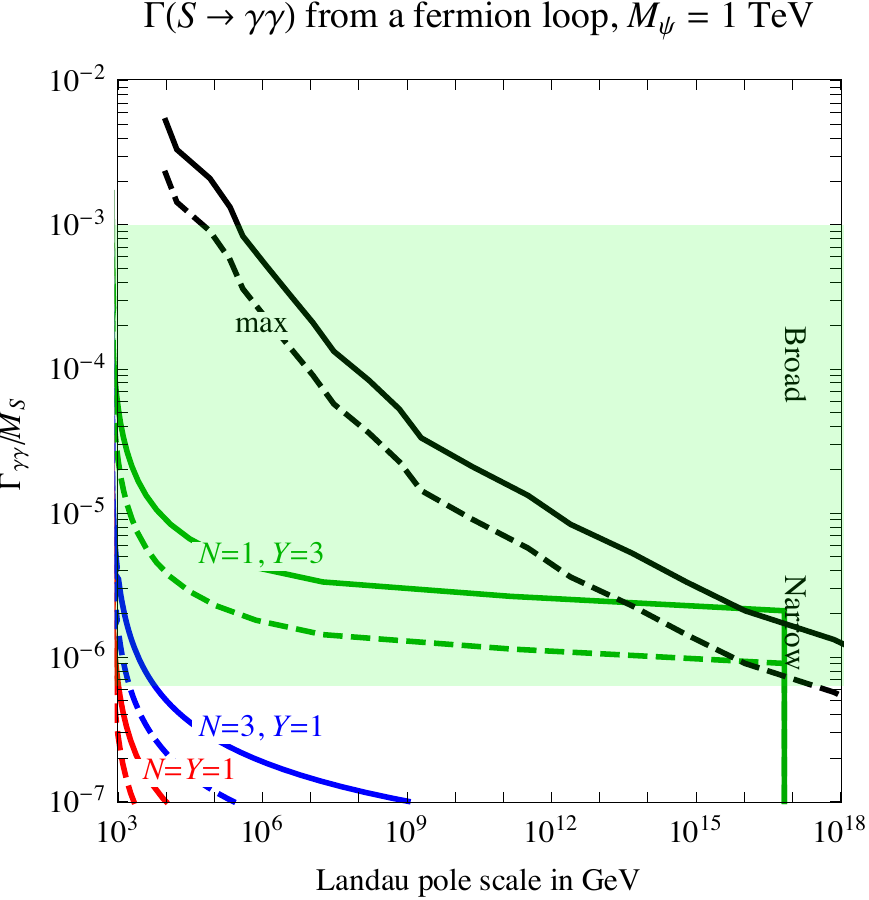}
$$
\caption{\em Maximal $\Ggg$ generated by a fermionic loop compatible with
perturbativity considering a $750\GeV$ scalar (dashed curves) or 
pseudo-scalar (continuous curves) with a CP-conserving Yukawa coupling.
The green band shows the value of $\Ggg$ favored by the $750 \GeV$ excess, assuming that $S$ has a narrow (lower) or broad (upper) width.
\label{fig:fermion}}
\end{center}
\end{figure}

\section{A fermionic loop}\label{psi}
We couple $S$ to $N$ fermions $\psi$ with mass $M_\psi$, hypercharge $Q=Y$
and singlet under $\SU(2)_L$.
We assume that the $N$ fermions have the same mass and same couplings,
such that the  Lagrangian 
\beq  \Lag = \Lag_{\rm SM} + \frac{(\partial_\mu S)^2}{2} +  \bar\psi(i\slashed{D}-M_\psi )\psi+
[\X \bar\psi (y + i \, \tilde y\gamma_5) \psi+\hbox{h.c.}]-
V(S) - V(S,H)
\label{eq:YukSQ}\eeq
respects a $\SU(N)$ symmetry: this choice simplifies computations and maximises $\Ggg$.
The potential is $V(S)=\frac{1}{2}M_S^2 S^2 + \lambda_S S^4$.
The Yukawa coupling $y$ ($\tilde y$) is present if $S$ is a scalar (pseudo-scalar).
$y$ and $\tilde{y}$ have the same RGE, and $\tilde y$ contributes more to $S\to\gamma\gamma$ than $y$ (see e.g.~\cite{750fits}).
If $S$ is a pseudo-scalar the loop function is maximal at $M_\psi=M_S/2$, giving
\beq\frac{\Gamma_{\gamma\gamma}}{M}\approx 0.6~10^{-6} N^2 \tilde y^2 Y^4.\eeq
Allowing $\SU(N)$ to become a gauge symmetry with gauge constant $g$, the relevant RGE are
\begin{eqnsystem}{sys:RGENf}
(4\pi)^2 \beta_{g_Y} &=& g_Y^3 (\frac{41}{6}+\frac{4N}{3} Y^2)\\
(4\pi)^2 \beta_g &=& -b g^3\qquad  b = \frac{11}{3}N - \frac{2}{3}- \cdots\\
(4\pi)^2 \beta_{y} &=& (2N+3)y^3 -y (6 g_Y^2 Y^2 + 3\frac{N^2-1}{N}  g^2) \label{eq:betay} \\
(4\pi)^2\beta_{\lambda_S} &=& 72\lambda_S^2 + 2N y^2 (4\lambda_S- y^2) \label{eq:betalambda}
\end{eqnsystem}
where  $\beta_\theta \equiv d\theta/d\ln\mu$ and
$\cdots$ denotes the contribution of extra possible particles charged under $\SU(N)$.
For simplicity, we assumed a vanishing quartic coupling $|S|^2 |H|^2$.\footnote{This coupling was considered
in~\cite{vac750} and helps in stabilising the electroweak vacuum~\cite{1203.0237}.}

\medskip

Assuming $g=0$ and ignoring eq.\eq{betalambda} we reproduce the results
of~\cite{Kam,Urb,750fits}, that we plot in
figure\fig{fermion} as the maximal value of $\Ggg$ as function of the Landau poles scale.
We have taken into account the RGE for $\lambda_S$, eq.\eq{betalambda},
which was partially considered in~\cite{vac750}.
We impose that $\lambda_S$ does not hit a Landau pole, and that it does not lead to too fast vacuum decay:
\beq -\frac{0.016}{1+0.01\ln \mu/M_S} < \lambda_S(\mu) \circa{<}4\pi . \label{eq:lambdaS}\eeq
A look at the RGE shows that the maximal $\Ggg$ is obtained
for small $N=1$ and for $Y$ as large as allowed by Landau poles for hypercharge,
which corresponds to uninteresting values $Y\sim 10$.
Thereby, we also plotted the maximal $\Ggg$ at fixed values of $Y$ and $N$.
We see that $\Ggg\circa{<} 10^{-6}$ can be obtained within models with reasonable $Y\sim 1$ and $N\circa{<}3$ that remain perturbative up to the Planck scale.
Larger values of $\Ggg$ need new non-perturbative physics not much above the TeV scale,
especially if the fermion $\psi$ is colored, such that it can also mediate $S\to gg$ but needs to be heavier
of about 1 TeV in view of LHC bounds.


\subsubsection*{Gauged SU($N$)}

Finally, we consider the new class of models obtained gauging $\SU(N)$.
Such models interpolate between weakly-coupled models ($g=0$)
and strongly-coupled models ($g$ becomes non perturbative around $M_S$)
considered in the literature~\cite{750fits,nonpert}.
It is interesting to notice that, even without considering the non-perturbative limit,
a perturbative $g$ allows to obtain qualitatively larger values of $y$ and thereby of $\Ggg$ without hitting Landau poles
than in the $g=0$ limit.
Indeed, if $g>0$, the RGE for $y$, eq.\eq{betay}, implies that the low energy value of $y$ is attracted towards the 
Pendleton-Ross infra-red fixed point~\cite{TAF}
\beq \frac{y^2}{g^2}\to
\frac{3(N-1/N)-b}{2N+3}
\label{eq:PR}
\eeq
provided that the latter term is positive, $b<3(N-1/N)$.
In such a case, $y$ at low energy can become arbitrarily large without hitting Landau poles,
given that the same holds for $g$.
For example, in the limit of large $N$ and small $b$ one has $y^2/g^2\to 3/2$.

If instead the latter term in eq.\eq{PR} is negative the infra-red fixed point does not exist, and
 adding a $g>0$ does not give
a result qualitatively different  from in the $g=0$ limit.

\section{A scalar loop}\label{min}
We now consider the scalar case, which requires discussing the (meta)stability of the full potential.
Thereby we first consider the case of a single scalar.

\subsection{A single charged scalar}\label{scal1}

We start considering the following minimal model, where the SM is extended by adding
a neutral real scalar  singlet $S$ and one
complex singlet $X$ with hypercharge $Y=Q$.
The scalar potential is
\beq V(H,S,X) = - \frac{M_h^2}{2} |H|^2 + \lambda_H |H|^4 + \lambda_{HS} |H|^2 S^2 + \lambda_{HX} |H|^2 |X|^2+
\kappa_{HS} S|H|^2+
 V(S,X) \eeq
where the terms involving only the new scalars $S$ and $X$ are
\beq
V (S,X)= \frac{M_S^2}{2} S^2 +M_X^2 |X|^2+\lambda_S S^4 + \lambda_{XS} S^2 |X|^2 +\lambda_X|X|^4+\frac{\kappa_S}{3} S^3 +\kappa_{XS} S|X|^2 . \label{eq:V}\eeq
At very large field values $S,X\gg M_S$ the potential is stable if $\lambda_S,\lambda_X>0$ and $\lambda_{XS}^2<4 
\lambda_S\lambda_S$.
The resulting $S$ width into photons is 
\beq \frac{ \Gamma(S\to\gamma\gamma)}{M}=
\frac{\alpha^2_{\rm em}}{256\pi^3}\bigg|
 \frac{\kappa_{XS}M_S}{2M_{X}^2} Q^2  F\left(\frac{4M_X^2}{M_S^2}\right)
\bigg|^2 
\label{eq:gamrat}\eeq
where the loop function $F$ is
\beq F(x) = x\bigg[ x \arctan^2\left(\frac{1}{\sqrt{x-1}}\right) -1\bigg]  \stackrel{x\to\infty}{=} \frac{1}{3}  .\eeq
Considering the potential as function of $S$ only, absolute stability is satisfied for
$ |\kappa_S|^2 < 18M_S^2 \lambda_S$.
In the presence of both $S$ and $X$, absolute stability can be again computed analytically, 
although the result is too long to be presented.
The main qualitative feature is that the
upper bound on $\Gamma(S\to\gamma\gamma)\propto|\kappa_{XS}|^2$ grows proportionally
to some combination linear in the quartics $\lambda_S,\lambda_{XS},\lambda_X$. 
This means that the scalar loop contribution to $S\to\gamma\gamma$ is limited by perturbativity of the quartics,
just like a fermion loop contribution is limited by perturbativity of the Yukawa $y^2$.
Then our goal is generalising to scalar case
the result found in the fermionic case and shown in fig.\fig{fermion}.

\begin{figure}[t]
\minipage{0.5\textwidth}
  \includegraphics[width=.9\linewidth]{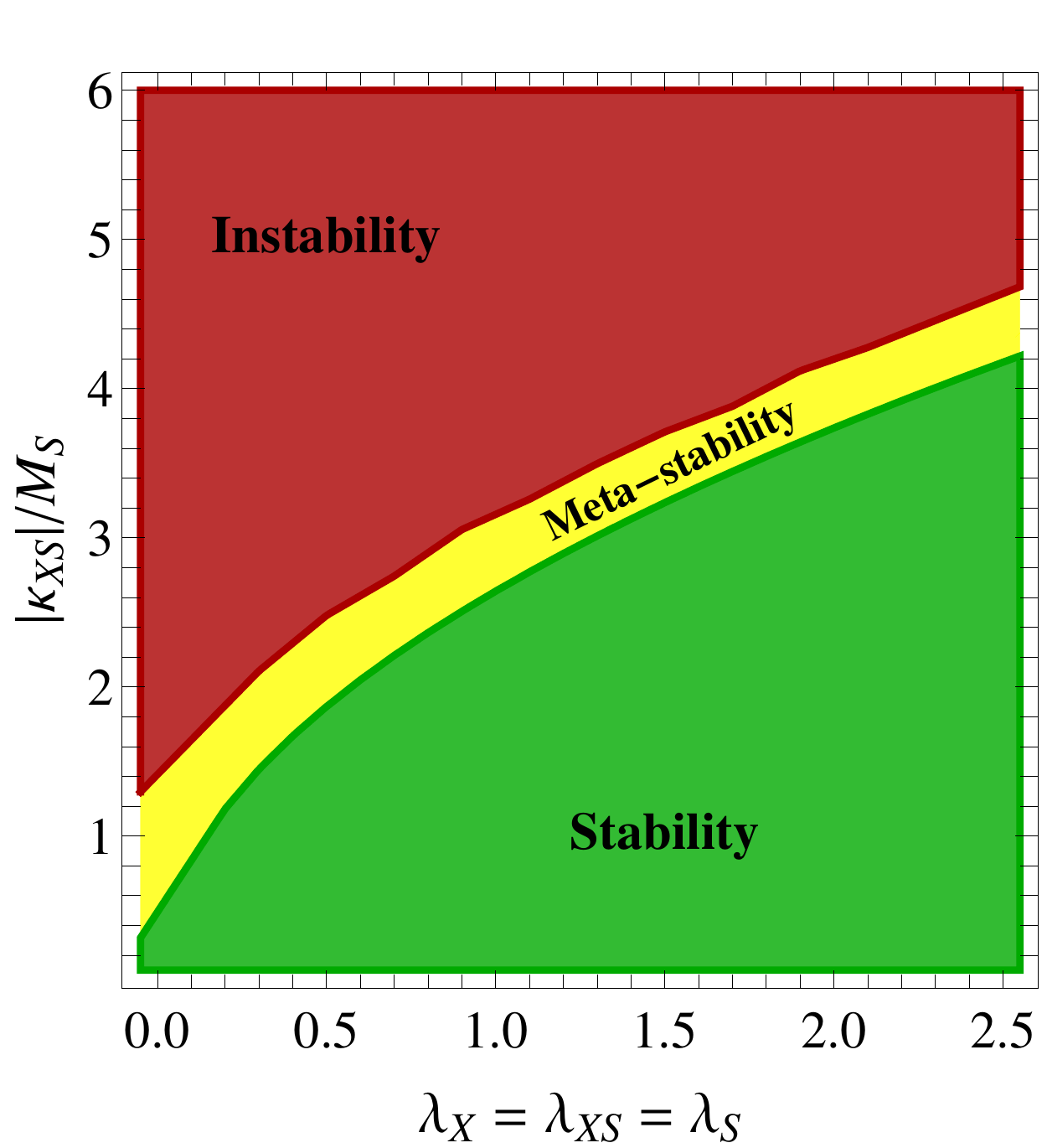}
\endminipage\hfill
\minipage{0.5\textwidth}
  \includegraphics[width=.9\linewidth]{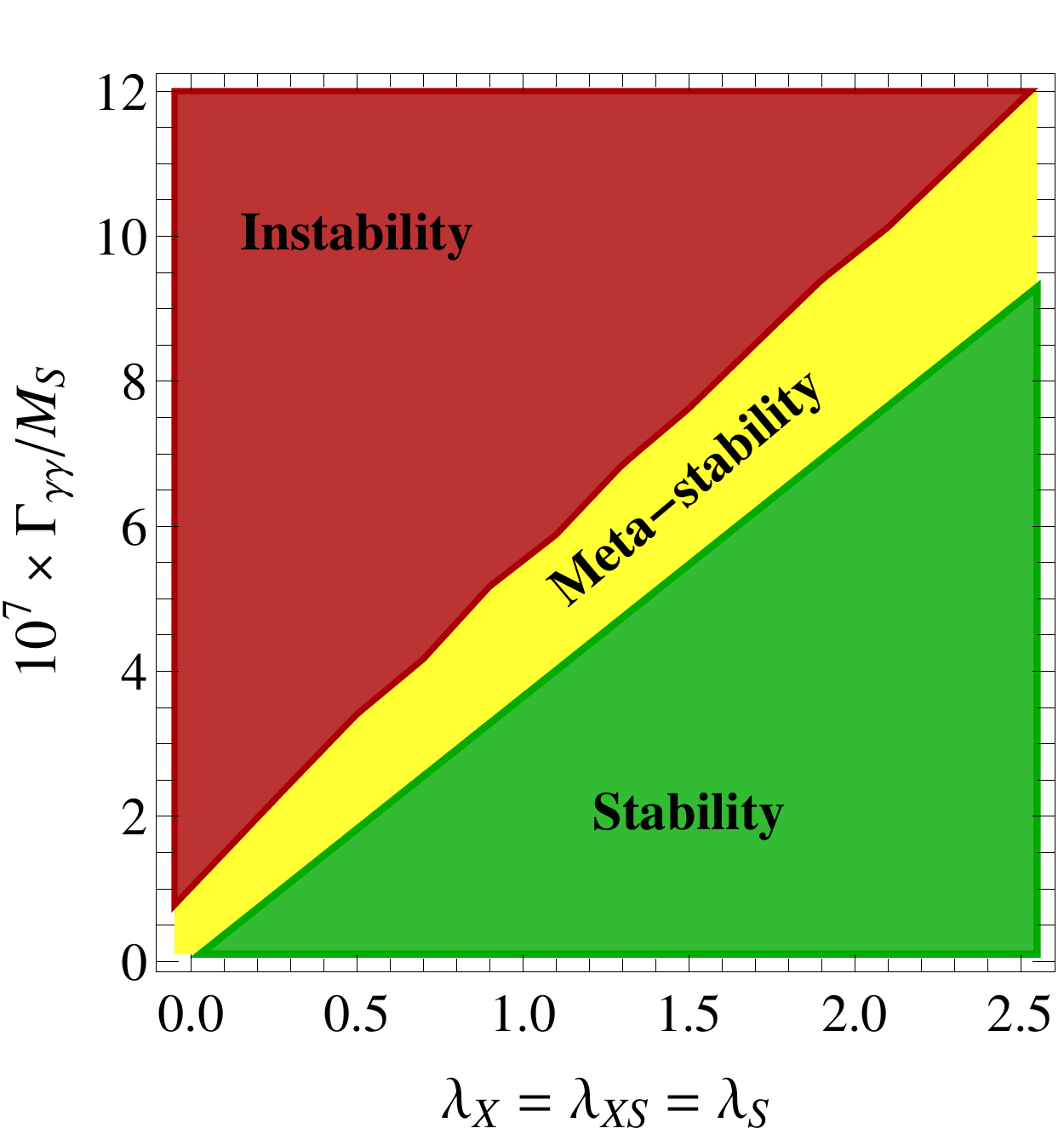}
\endminipage
\caption{\em 
{\bf  Left (a):} Maximal cubic $|\kappa_{XS}|/M_S$ 
allowed by stability (green) and by meta-stability (yellow)
as function of $\lambda_X=\lambda_S=\lambda_{XS}$.
{\bf  Right (b):} The corresponding value of $\Gamma_{\gamma\gamma}/M$
assuming that the scalar $X$ has charge $Q=1$.
Vacuum decay is too fast in the red regions.
\label{fig:scalarsample}}
\end{figure}

\subsubsection{Meta-stability}
The meta-stability condition can be computed only numerically, and is weaker than the stability condition,
altought they are qualitatively similar.

For the numerical computation we use the tool-chain {\tt SARAH}--{\tt SPheno}--{\tt Vevacious}: we implemented the minimal model of eq.\eq{V}
in {\tt SARAH}  \cite{Staub:2008uz,Staub:2009bi,Staub:2010jh,Staub:2012pb,Staub:2013tta,Staub:2015kfa} and generated the {Fortran code} for {\tt SPheno} \cite{Porod:2003um,Porod:2011nf} to get a spectrum generator for the model. {\tt SPheno} was used to compute all masses and branching ratios, and the produced spectrum file is then given to {\tt Vevacious} \cite{Camargo-Molina:2013qva} as input to  check the stability of the electroweak vacuum. For this purpose, we generated a model file with {\tt SARAH} for  {\tt Vevacious} which includes the possibility of VEVs for the charged scalar beside to ones for the neutral states. {\tt Vevacious} checks the stability of the scalar potential via a  homotopy method which guarantees to find all minima of the tree-level potential. In the case that there is a  minimum deeper than the desired one, it calls {\tt ComsoTransitions}  \cite{Wainwright:2011kj} to calculate the life-time of the vacuum. 
The decay rate $\Gamma$ per unit volume for false
 vacuum decay can be written as \cite{Coleman:1977py, Callan:1977pt} 
\begin{equation}
\frac{d\wp}{dV dt} = \frac{ e^{-S} }{R^4}
\label{eq:tunneling_time}
\end{equation}
where $R\approx 1/M_S$ is the size of the bounce and
$S$ is the action of the bounce. At tree level it is given by
\begin{equation}
S = \int d^4 x\left(\frac{(\partial_\mu S)^2}{2} + |D_\mu X|^2+ V(S,X)\right). 
\end{equation}
{\tt CosmoTransitions} finds the multi-field optimal `path' to tunnel from the false to the 
true vacuum using the B-splines 
algorithm. For more technical details we refer to  \cite{Wainwright:2011kj}.

Integrating over our past-light cone, taking into account the expansion of the universe, we find
the present value of the vacuum-decay probability $\wp$,
\beq
\wp_0 = 0.15~ \frac{e^{-S}}{(RH_0)^4} ,
\label{eq:probdoggi}
\eeq
where  $H_0\approx 67.4\,{\rm km/sec~Mpc}$ is the present Hubble rate.
A probability $\wp_0$ larger than $10\%$ is obtained for $S>412$.
Based on the result of this calculation, we label as meta-stable a point such that $\wp_0>10\%$,
and unstable otherwise.


To start and to illustrate the result,
we consider the special case $\lambda_X = \lambda_{XS}=\lambda_S$ and
$\kappa_S=0$.
Furthermore we fix $M_X=M_S/2$, which is the value that maximises $\Ggg$.
Fig.\fig{scalarsample}a  shows the resulting stability region (green):
the maximal $|\kappa_{XS}|$ grows proportionally to the squared root of the couplings.
The extra region allowed by meta-stability  (in yellow) has a similar shape.
In fig.\fig{scalarsample}a we show the corresponding $\Ggg$ rate, assuming
a single scalar $X$ with $Q=1$: we see that a phenomenologically relevant value $\Ggg\circa{>}10^{-6}$
needs quartic couplings of order 1.

\subsubsection{Perturbativity limits}
In order to quantify if a TeV-scale value of the quartics is `too large',
we solve their one-loop renormalisation group equations
and compute the RGE energy scale $\mu=\Lambda$ at which a coupling hits a Landau pole.
The  RGEs that involve only the quartic couplings of $S,X$ are (the full set of RGE is given later)
\begin{eqnsystem}{sys:RGEmain}
\beta_{\lambda_X}&=& \frac{1}{(4\pi)^2 }\left[
20 \lambda_X^2 + 2\lambda_{XS}^2\right],\\
\beta_{\lambda_{XS}}  &=&  \frac{1}{(4\pi)^2 }\left[ 8\lambda_{XS}^2+8 \lambda_{XS} (\lambda_X+3\lambda_S)
\right],\\
\beta_{\lambda_S}&=&  \frac{1}{(4\pi)^2 }\left[  \lambda_{XS}^2 + 72\lambda_{S}^2\right].
\end{eqnsystem}
A large coupling leads to a Landau pole at low energy; in such a case
$\Lambda$ can be approximated as
\beq 
\Lambda \approx M_S \exp\min_\lambda \frac{\lambda}{\beta_\lambda}\eeq
which becomes exact in the case of a single quartic coupling.

\medskip

\begin{figure}[t]
\begin{center}
$$\includegraphics[width=0.45\textwidth]{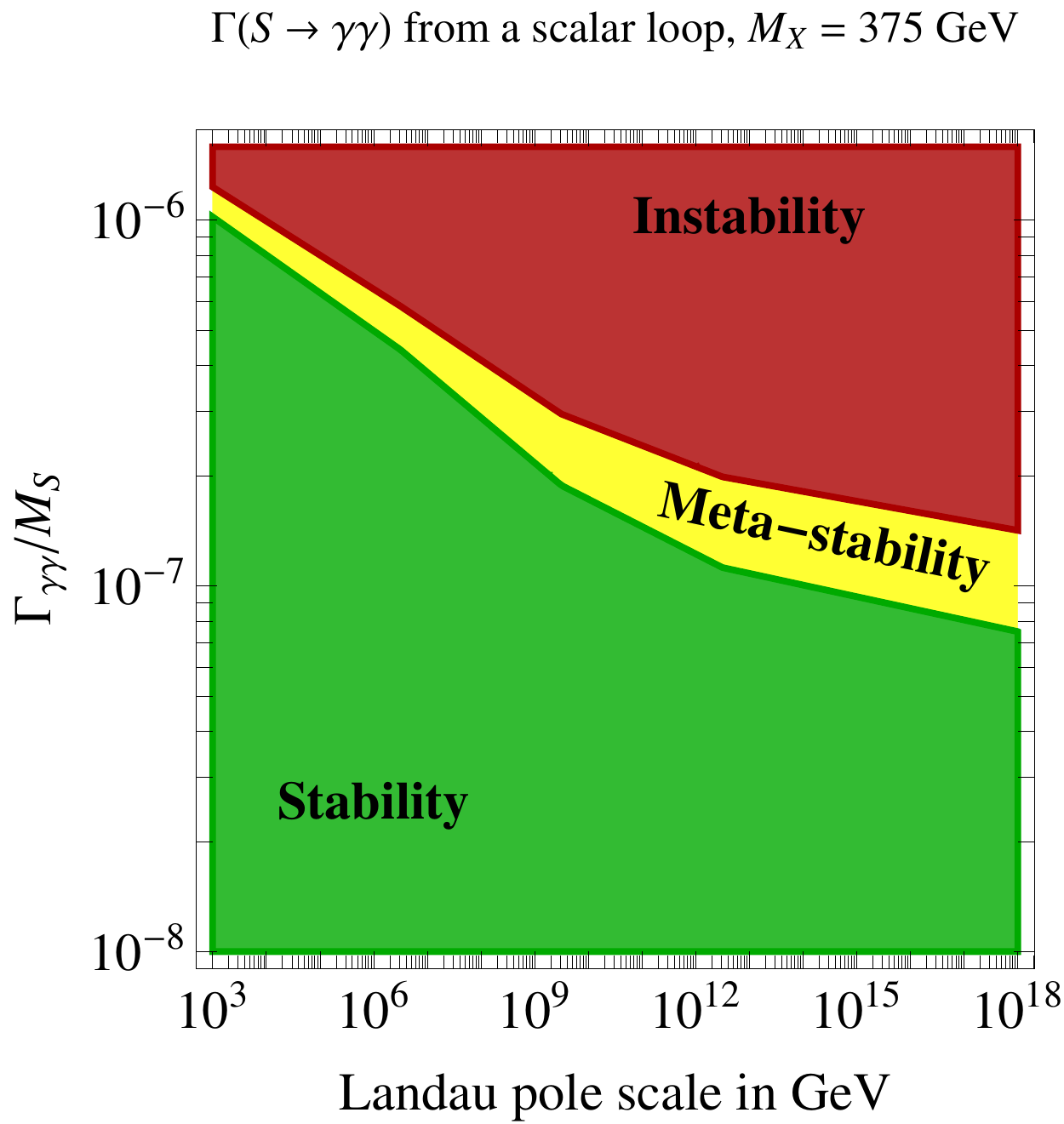}\qquad
\includegraphics[width=0.46\textwidth]{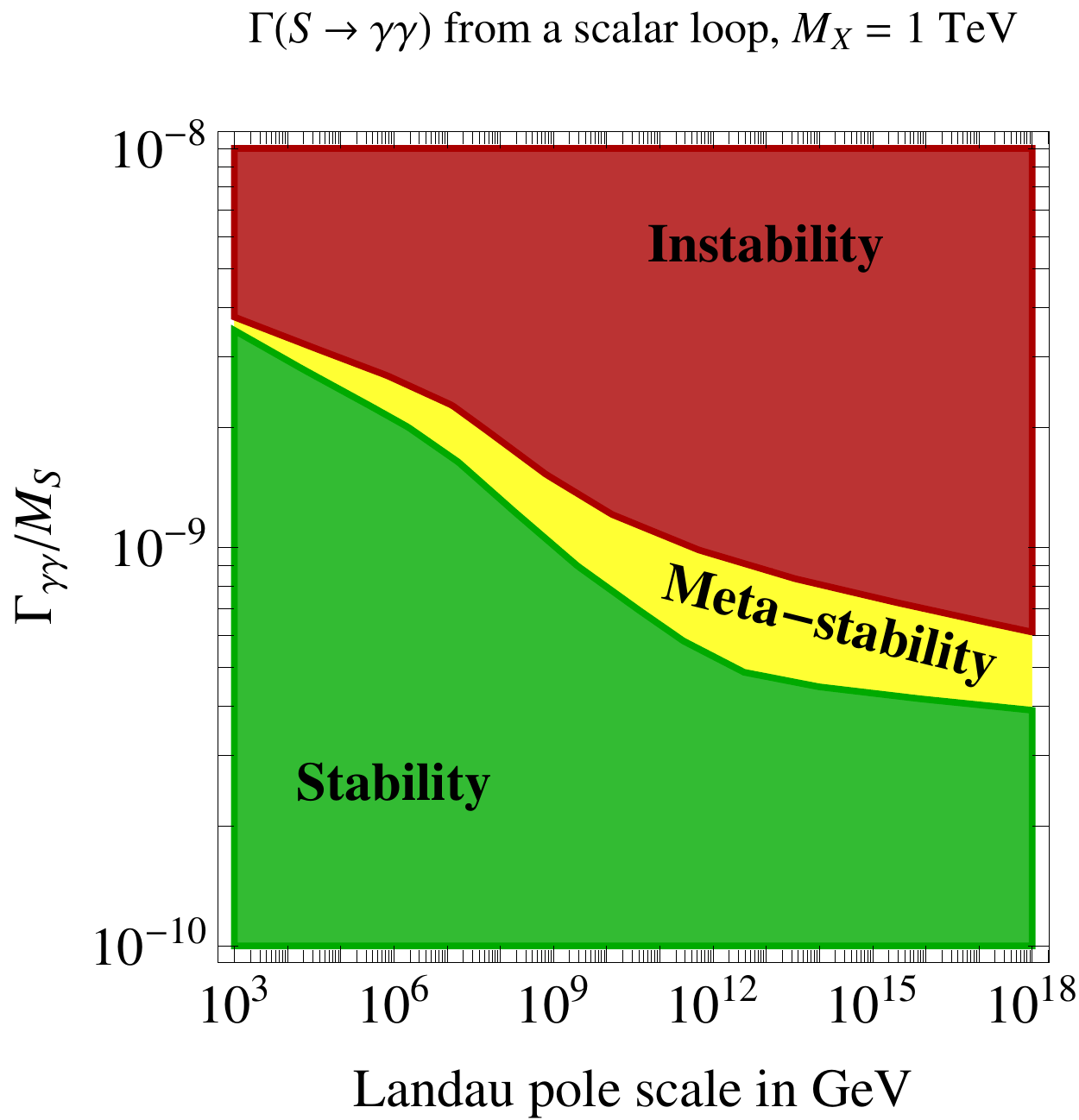}
$$
\caption{\em 
Maximal $\Ggg$ allowed by perturbativity considering a scalar $S$ with a cubic coupling
to one singlet charged scalar $X$ with $Q=Y=1$.
\label{fig:scalargeneric1}}
\end{center}
\end{figure}

We next perform a full scanning,
picking random points in the parameter space of the model,
and checking if stability and/or meta-stability are satisfied;
in such a case we compute  $\Ggg$ and the Landau pole scale.
The final result is shown in fig.\fig{scalargeneric1}:
like in the fermionic case, a larger $\Ggg$ implies a Landau pole at lower energy.
Actually, the maximal $\Ggg$ is a factor of few lower than in the corresponding  fermionic case.

The $\Ggg$ width can be increased by allowing for a scalar $X$ with bigger charge
and/or for multiple states $X$.
However, these possibilities are limited by Landau poles for the hypercharge gauge coupling 
and by precision data, as studied in the next section.

\begin{figure}[t]
\begin{center}
$$\includegraphics[width=0.45\textwidth]{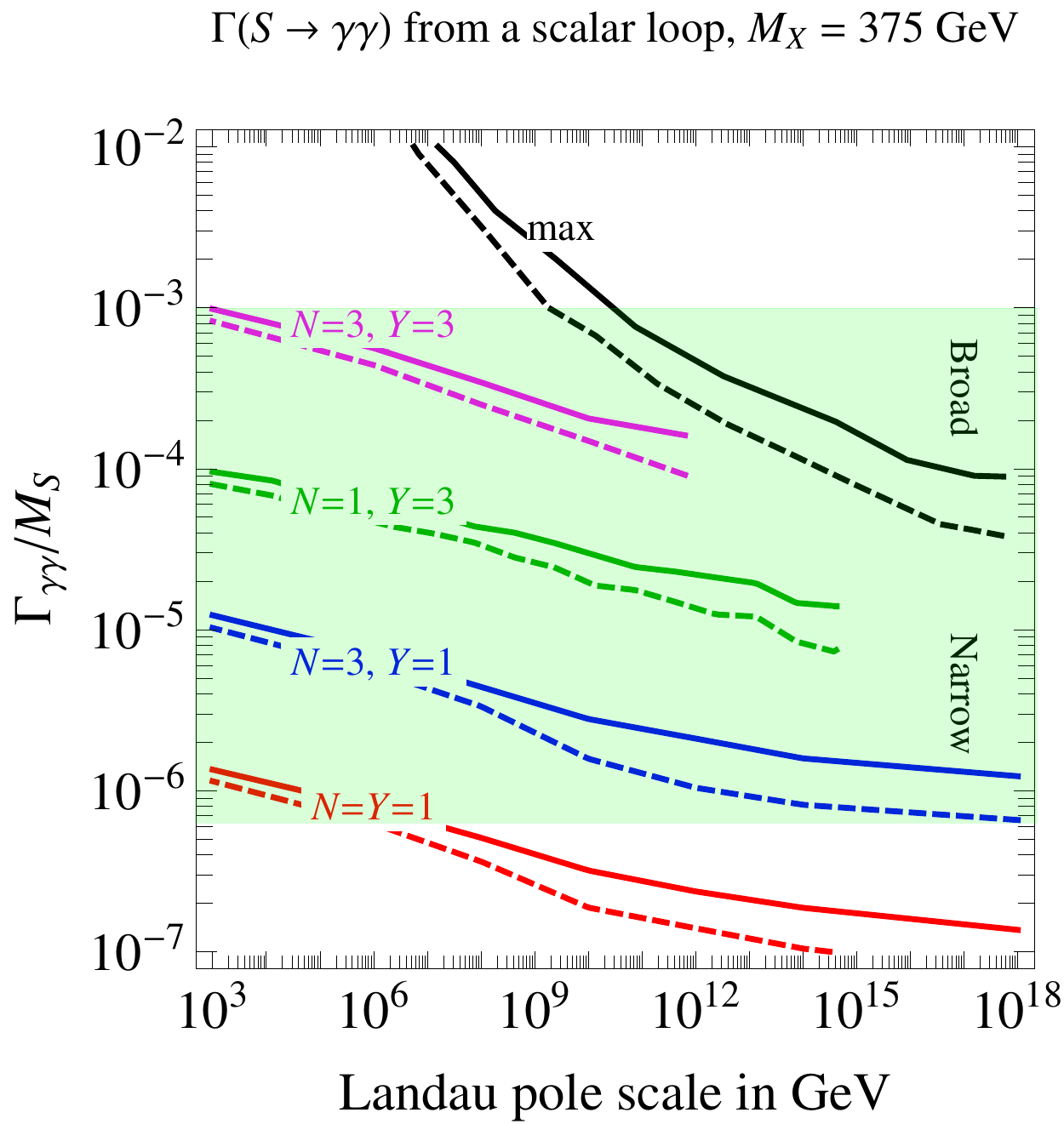}\qquad
\includegraphics[width=0.45\textwidth]{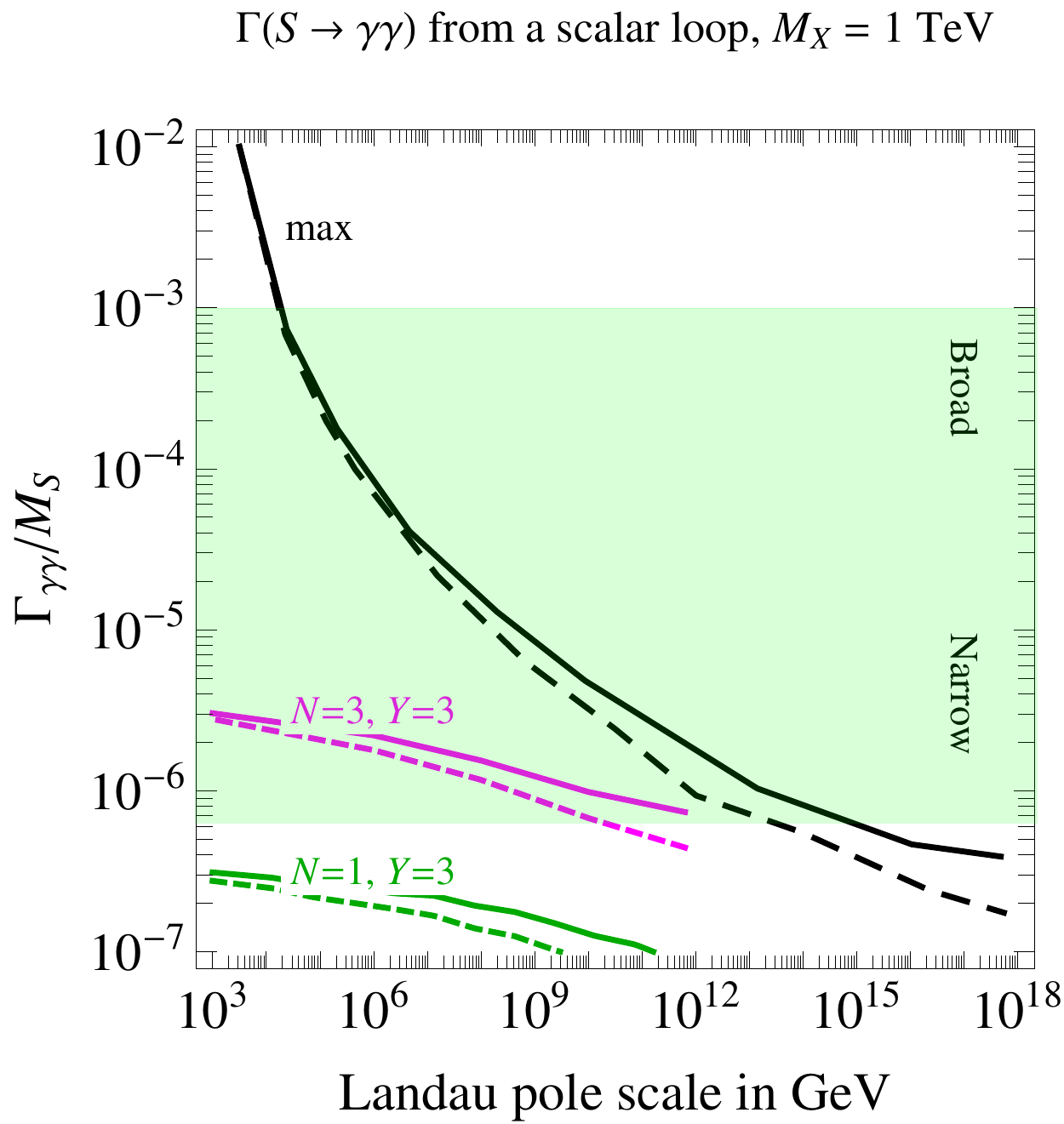}
$$
\caption{\em
Maximal $\Ggg$ generated by a scalar loop compatibly with 
vacuum stability (dashed curves) or by meta-stability (continuous curves)
as function of the scale at which the theory becomes non-perturbative.
The upper curves in black refer to a generic set of scalars;
the lower curves to some special case: a single scalar $(N=1)$ with unity hypercharge $(Y=1)$,
multiple fields (blue, $N=3$),  bigger hypercharge (green, $Y=3$) and both (magenta, $N=Y=3$).
The maximal $\Ggg$ is obtained for $M_X=M_S/2$ (left panel);
in the right panel we consider $M_X=1\TeV$, which is allowed by LHC data if the scalar fields are colored.
\label{fig:scalargeneric}}
\end{center}
\end{figure}

\subsection{Many scalars}\label{many}
We now generalize the results of the previous section including more charged scalars.
We consider  $N$ scalars $X$ with hypercharge $Y$
and singlet under SU(2)$_L$, 
assumed to lie in a fundamental representation of an extra SU($N$) global or gauge symmetry.
This means that all scalars have the same mass and the same cubic: this choice maximises their effect on $\Ggg$.
The vacuum stability and meta-stability conditions remain the same as in the previous section: we just need to 
take into account the enhancement in $\Ggg$ and the modified perturbativity conditions.

\subsubsection{Perturbativity limits}

We write the RGE including all relevant SM couplings: the gauge couplings $g_3$, $g_2$ and $g_1\equiv \sqrt{5/3}g_Y$, the top 
Yukawa coupling $y_t$ and the quartic couplings $\lambda_H$, $\lambda_{HS}$ and $\lambda_{HX}$.
We also consider the (possibly vanishing) gauge coupling of $\SU(N)$, $g$.
The RGEs for the dimensionless couplings $\lambda_X$, $\lambda_{XS}$, $\lambda_S$, $g_1$ and $g$  are 
\begin{eqnsystem}{sys:RGEN}
(4\pi)^2 \beta_{\lambda_X} &=&   4(N+4) \lambda_X^2 + 2\lambda_{XS}^2-\frac{36Y^2 g_1^2\lambda_X}{5} +\frac{54 Y^4 }{25}g_1^4+ 2\lambda_{HX}^2+ \nonumber\\
&&+ \frac{3(N-1)(N^2+2N-2)}{4N^2} g^4 - \frac{6(N^2-1)}{N} g^2 \lambda_X, \\
(4\pi)^2 \beta_{\lambda_{XS}} &=&  8\lambda_{XS}^2+4 \lambda_{XS} \left[(1+N)\lambda_X+6\lambda_S-\frac{9Y^2 g_1^2}{10} \right]+\nonumber\\
&&+4\lambda_{HS}\lambda_{HX} - \frac{3(N^2-1)}{N} \lambda_{XS} g^2\\
(4\pi)^2 \beta_{\lambda_S} &=&   N \lambda_{XS}^2 + 72\lambda_{S}^2+2\lambda_{HS}^2, \\ 
(4\pi)^2 \beta_{g_1} &=& g_1^3\frac{41+2NY^2}{10}, \\
   (4\pi)^2 \beta_{g}&=&  g^3 \left( -\frac{11}{3} N +\frac16\right).
\end{eqnsystem}
We included the quartic couplings $\lambda_{HS}$ and $\lambda_{HX}$
that involve the Higgs boson because, although they negligibly affect the
non-perturbativity issue, they unavoidably enter into the RGEs for $\lambda_X$, $\lambda_{XS}$, $\lambda_S$.
Indeed the quartic $\lambda_{HX}$ is unavoidably generated by hypercharge interactions
because both $H$ and $X$ are charged;
then a $\lambda_{HS}$ coupling is generated too as dictated by the following RGEs:
\begin{eqnsystem}{sys:RGENH} 
(4\pi)^2 \beta_{\lambda_{HX}} &=& \lambda_{HX} \left[4(1+N) \lambda_X-\frac{(36 Y^2+9)g_1^2 }{10}-\frac{9 g_2^2}{2}+12 \lambda_H+6 y_t^2\right]\nonumber\\ && + 4 \lambda_{HS} \lambda_{XS}+4 \lambda_{HX}^2 + \frac{27 g_1^4 Y^2}{25} - \frac{3(N^2-1)}{N} \lambda_{HX} g^2,\\ 
(4\pi)^2 \beta_{\lambda_{HS}} &=& 2 N \lambda_{XS} \lambda_{HX} +8 \lambda_{HS}^2 +\lambda_{HS}\left(24\lambda_S-\frac{9g_1^2}{10}-\frac{9g_2^2}{2}+6y_t^2+12 \lambda_H\right),\\
(4\pi)^2 \beta_{\lambda_{H}} &=&2 \lambda_{HS}^2+N \lambda_{HX}^2+ \frac{27 g_1^4}{200}+\frac{9 g_1^2 g_2^2}{20}+\frac{9 g_2^4}{8}+\lambda_{H} \left(-\frac{9 g_{1}^2}{5}-9 g_2^2+12 y_t^2\right)\nonumber \\ &&+24 \lambda_H^2-6 y_t^4,\\
(4\pi)^2 \beta_{y_t} &=&y_t\left(\frac92 y_t^2-\frac{17g_1^2}{20} -8g_3^2-\frac{9g_2^2}{4}\right),\\
(4\pi)^2 \beta_{g_2}&=&-\frac{19 g_2^3}{6},\qquad   (4\pi)^2 \beta_{g_3}=- 7 g_3^3.
\end{eqnsystem}
We are now ready to present our final result. 
Setting $g=0$ (global $\SU(N)$ symmetry)
in figure~\fig{scalargeneric}b we show the maximal value of $\Ggg$, as function of the scale at which a
Landau pole develops.
$\Ggg$ gets significantly enhanced, even by orders of magnitude,
with respect to the minimal case $N=Y=1$ considered in  section~\ref{scal1}.
The plot also shows the special cases $N=3$ and $Q=3$.
The final result is similar to the analogous fermionic result, shown in fig.\fig{fermion}.

%
%

%

\subsubsection{Gauged SU($N$)}
The gauging of the $\SU(N)$ symmetry allows, both in the fermionic and in the 
scalar case, to get larger values of $\Ggg$ without hitting Landau poles.
Indeed, if $g$ runs becoming larger at low energy, 
the quartic $\lambda_X$ gets driven to comparably large values, being
attracted towards the quasi-fixed point~\cite{TAF}
\beq 
\frac{\lambda_X}{g^2} \to \frac{s_{\lambda g} -b+  \sqrt{(s_{\lambda g}-b)^2-4 s_\lambda s_g}}{2 s_\lambda  } 
\label{irattra}
\eeq
where  $b,s_g,s_\lambda,s_{\lambda g}$ are constants that parameterise the RGE coefficients as
\beq 
(4\pi)^2\beta_g = -b g^3\ ,\qquad
(4\pi)^2 \beta_{\lambda_X} =2[  s_\lambda  \lambda_X^2  - s_{\lambda g} \lambda_X  g^2 + s_g  g^4] \ ,
\label{RGElambdaX}\eeq
For example we find
$\lambda_X/g^2\to (3+\sqrt{6}) /4$
in the limit of large $N$ and small $b\ll s_{\lambda g}$.
Like in the fermionic case,
the qualitative properties of $\lambda_X$ from eq.s~(\ref{RGElambdaX}) depend on the sign of 
\be E\equiv \frac{(s_{\lambda g}-b)^2-4 s_\lambda s_g}{4s_g^2}\ee
(see the last article in~\cite{TAF}). 
The infra-red fixed point exists if $E\geq 0$; in such a case
there is no Landau pole for values of the quartic such that $\sqrt{E}\leq D\leq \Lambda_0+\sqrt{E}$,
where 
\be D\equiv \frac{s_{\lambda g}-b}{2 s_g}\ ,\qquad \Lambda_0 \equiv \frac{g^2(\mu_0)}{\lambda(\mu_0)} \ee
and $\mu_0$ is some reference energy. 
If instead $E<0$ there is always a Landau pole.

\begin{figure}[t]
\begin{center}
$$\includegraphics[width=0.45\textwidth]{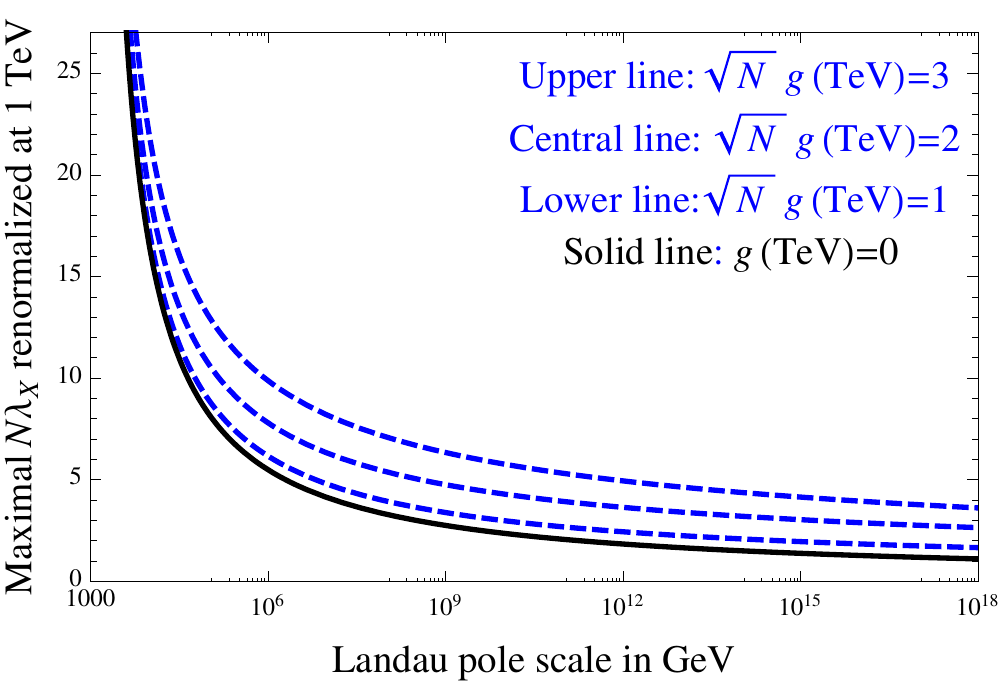}\qquad
\includegraphics[width=0.45\textwidth]{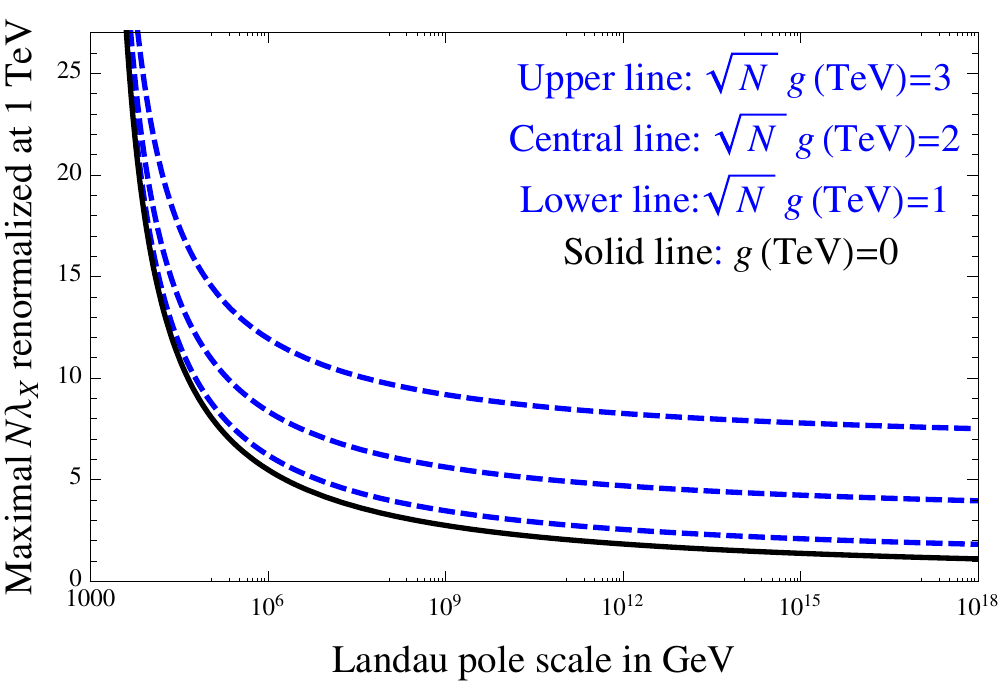}
$$
\caption{\em
Maximal $N \lambda_X$ at low energy
(chosen to be 1 TeV) as function of the maximal energy at which the theory holds without hitting Landau poles.
We consider the large $N$ limit and fixed values of the `t Hooft coupling $\sqrt{N} g$ at $\TeV$ energy.
 {\bf Left:} no fixed points.  {\bf Right:} the gauge beta function is reduced to $b=N$ such that
 a fixed point for $y$ arises.
\label{fig:NlambdaLambdaLP}}
\end{center}
\end{figure}

This situation is illustrated in fig.~\ref{fig:NlambdaLambdaLP} and in its caption. 
The plot on the right has $E\geq 0$ (so a fixed point is allowed),
and the low-energy value of $y$ remains finite even assuming no Landau pole up to arbitrarily large energy.
The plot on the left has $E<0$ (no fixed point), and $y$ at low energy
must be small if the theory cannot have Landau poles up to higher energy:
allowing a $g\neq 0$ only has a minor effect with respect to the $g=0$ limit.

%
%

\begin{figure}[t]
\begin{center}
$$\includegraphics[width=0.45\textwidth]{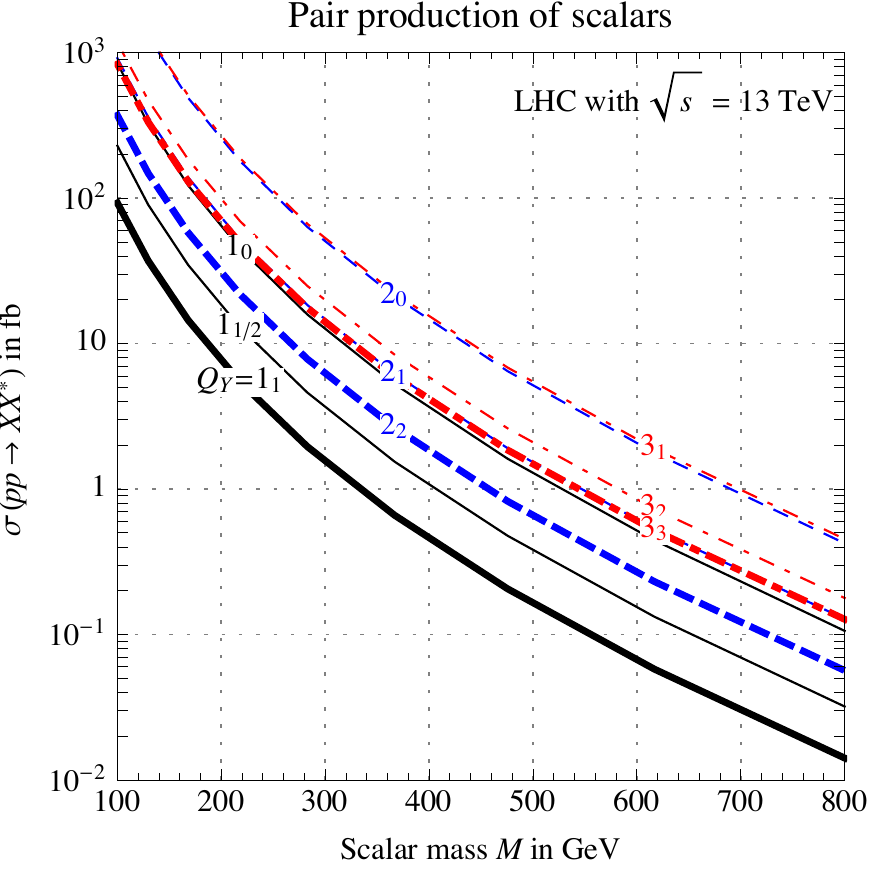}\qquad\includegraphics[width=0.45\textwidth]{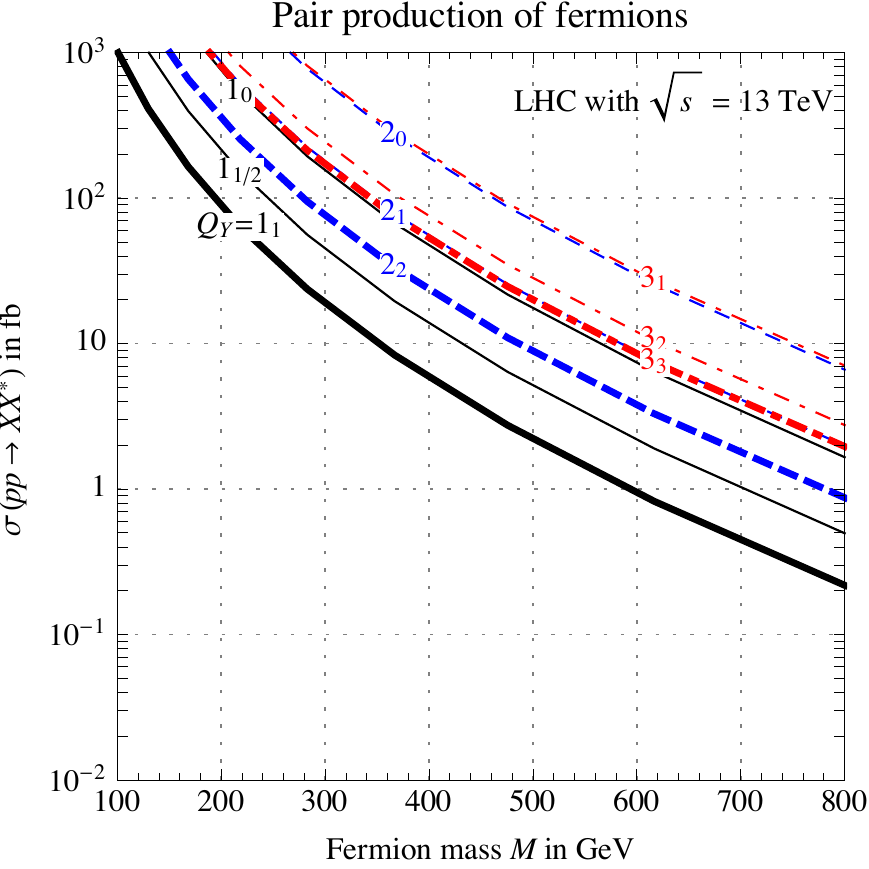}$$
\caption{\em Cross section $pp\to XX^*$ for producing two uncolored particles (scalars in the left plot and fermions in the right plot)
 with charge $Q$ and
hypercharege $Y$, indicated as $Q_Y$.
\label{fig:LHC}}
\end{center}
\end{figure}

\section{Collider probes and Dark Matter}\label{col}
\subsection{Collider probes}

We now discuss how the scenario can be probed at colliders.
The partonic cross section 
$q_1 \bar q_2 \to X_1 \bar X_2$ for
pair production of two uncolored scalars or fermions $X_1 \bar X_2$ is 
\beq
\frac{d\sigma}{d\hat t}
=\frac{V_L^2 + V_R^2} {144\pi \hat s^2} \times \left\{
\begin{array}{ll}
  (2M_1^2 M_2^2 \!+\! \hat s^2 \!-\! 2(M_1^2+M_2^2)\hat{t}\!+\!2 \hat t^2 \!+\! \hat {s}(2\hat{t}-(M_1\!-\!M_2)^2)  & \hbox{fermion}\\
(M_1^2 M_2^2  - (M_1^2+M_2^2)\hat{t}+ \hat t^2 + \hat {s}\hat{t} ) &\hbox{scalar}
\end{array}\right.
\eeq
where
\beq V_A^2=\left\{
\begin{array}{ll}
 \displaystyle 
3 \bigg(Q_q Q_X \frac{ e^2}{\hat s}+
g_{q_A} g_X\frac{g_2^2/c_{\rm W}^2}{\hat{s} - M_Z^2}\bigg)^2
\qquad& \hbox{for $q\bar q\to X X^*$ }\\
3w_X\displaystyle\bigg(\frac{g_2^2}{\hat s-M_W^2}\bigg)^2 & \hbox{for $u\bar d\to X_1X_2$ }
\end{array}\right.\eeq
and  $g = T_3 - s_W^2 Q$ is the $Z$ coupling,
$A=\{L,R\}$.
So far we considered the case of a SU(2)$_L$ singlet: in such a case one has $w_X=0$.
Otherwise $w\neq 0$ if $A=L$ and $T_3(X_1) - T_3(X_2)=\pm1$:
$w_X=1$ if $X$ is a weak doublet;
$w_X=2$ if  $X$ is a weak triplet.
The resulting $pp$ cross section is plotted in fig.~\ref{fig:LHC} and grows as $NQ^2$.
As well known, the cross section for pair production of scalars (left) is $p$-wave suppressed and
about one order of magnitude smaller than the fermion pair production cross section (right).

\medskip

The experimental bounds on such cross sections depend on how $X$ decays.
A large variety of possibility exists; furthermore
gauged $\SU(N)$ could lead to `quirk' phenomena~\cite{quirk}.
Heavy leptons tend to give easily detectable signals,
potentially giving limits as strong as the present inverse luminosity $L$,
$\sigma\circa{<} \hbox{few}/L \approx  \fb$.
In such a case, fig.~\ref{fig:LHC} implies that new fermions and (to a lesser extend) new scalars
with large multiplicities and/or large charges are already excluded, if their masses
are around few hundred GeV.

\subsection{Dark Matter}

It is interesting to consider the case where $X$ lies in a SU(2)$_L$ multiplet that contains, as lightest component,
a neutral state that can be a Dark Matter candidate.
At colliders Dark Matter can be seen as missing energy. 
If the SU(2)$_L$ multiplet is quasi-degenerate (Minimal Dark Matter limit~\cite{MDM}), the decay products that allow to tag the event 
become soft and can be missed.
One needs to rely on initial state radiation, which can give an extra jet or photon or $Z$,
but with a smaller cross section, such that the signal can easily be below the SM backgrounds.
In this situation a large multiplicity of light $X$ particles ($M\circa{<}M_S/2$) becomes allowed.

The thermal freeze-out cosmological Dark Matter abundance is reproduced when the
$s$-wave DM (co)annihilation cross-section equals to $\sigma_0 \approx{1}/{(22\TeV)^2}$.
 In the Minimal Dark Matter limit the $\sigma_0$  induced by SM
gauge interactions  is given by
\beq \sigma_0 = \frac{\sum_R d_R^2 \sigma_{0}(R)}{(\sum_R d_R)^2}\eeq
where~\cite{MDM}
\beq  \sigma_{0}(n,Y) =\left\{\begin{array}{ll}\displaystyle
\frac{g_2^4\ (2n^4+17n^2-19) + 4  Y^2 g_Y^4 (41+8Y^2)+16g_2^2 g_Y^2 Y^2 (n^2-1)}{1024\pi nc M_{R}^2} & \hbox{fermion}\\
\displaystyle
\frac{g_2^4\ (3-4n^2+n^4) + 16\ Y^4 g_Y^4 + 8g_2^2 g_Y^2 Y^2 (n^2-1)}{256 \pi c n M_R^2}& \hbox{scalar}
 \end{array}\right.
\eeq
We have considered a multiple set of MDM representations $R$
that fill a $n$-dimensional representation of $\SU(2)_L$ with hypercharge $Y$.
Their number of degrees of freedom is $d_R=2cn $ (scalar) or
$d_R=4cn$ (fermion) where
 $c=1/2~(1)$ for a real (complex) representation.

Taking into account that extra annihilations mediated by $S$ are typically subdominant~\cite{750fits},
$N$ degenerate scalar doublets with $Y=1/2$ reproduce 
the observed DM abundance if their mass is $M = 540\GeV/\sqrt{N}$, which is lighter than
$M_S/2$ (providing decay channels for $S$) for $N\ge 2$.
Such doublets predict extra decays $S\to \gamma Z,ZZ,W^+W^-$ at an acceptable level~\cite{750fits}.
This shows that a consistent scenario can be obtained.
On the other hand, fermionic doublets or higher $\SU(2)_L$ multiplets such as triplets
cannot reproduce the DM abundance unless they have a very large multiplicity $N$.

If $\SU(N)$ is gauged its vectors could form quasi-stable Dark Matter~\cite{SUN}.

\subsection{Precision observables}

Given that it is difficult to directly detect quasi-degenerate Dark Matter weak multiplets  at LHC,
it is interesting to explore how they indirectly affect precision data.

New scalars or fermions with hypercharge $Y$ and mass $M_X$ 
belong to the class of `universal new physics' that affects precision data 
measurable at colliders with energy $\sqrt{s}\ll M_X$
only trough the ${\cal S,T,W,Y}$ parameters~\cite{STWY}.  
We assume that these particles are not coupled to the SM Higgs doublet, 
so that the ${\cal S}$ and ${\cal T}$ parameters are not affected.
On the other hand,  the ${\cal Y}$ and ${\cal W}$ parameters receive the following contributions~\cite{Shap} 
\begin{eqnsystem}{sys:WY} 
{\cal Y}&=&\sum_s \Delta b_Y^{(s)}\frac{\alpha_Y}{40 \pi}   \frac{M_W^2}{M_{s}^2}
+\sum_f \Delta b_Y^{(f)}\frac{\alpha_Y}{20 \pi}   \frac{M_W^2}{M_{f}^2},\\
{\cal W}
&=&\sum_s \Delta b_2^{(s)}\frac{\alpha_2}{40 \pi}   \frac{M_W^2}{M_{s}^2}
+\sum_f \Delta b_2^{(f)}\frac{\alpha_2}{20 \pi}   \frac{M_W^2}{M_{f}^2},
\end{eqnsystem}
%
where $\Delta b_Y^{(s)} = d_R Y^2/6$ and $\Delta b_Y^{(f)}= d_R Y^2/3$
are the usual contributions to the hypercharge beta-function coefficients coming from each $s$calar and $f$ermion.
$\Delta b_2^{(s)}$ and $\Delta b_2^{(f)}$ are the analogous coefficients for  the $\SU(2)_L$  beta functions.

The present experimental bound, $|{\cal Y}|\circa{<}2~10^{-3}$~\cite{STWY}, implies
 $\sum_R d_R Y^2\circa{<}1500 (M_R/375\GeV)^2$, which is too weak to have
a significant impact on our present analysis.
Comparable limits on this kind of effects can be obtained from the differential $pp\to \mu^+\mu^-$ 
cross section at LHC at large invariant mass~\cite{1410.6810}.

A future circular collider operating at the $Z$ peak can measure
${\cal W,Y}$ with improved accuracy. 
According to~\cite{FCCee}, 
a precision of $10^{-6}$ on $\sin^2\theta_W$ 
(the effective mixing angle defined trough $Z$ couplings)
is a reasonable goal.
The theoretical uncertainty can be brought down to the same level, expect 
for the uncertainty  coming from $\alpha_{\rm em}(M_Z)$, which presently is $18~10^{-6}$~\cite{FCCee}.
In any case this would be the dominant constraint on ${\cal Y}$ and ${\cal W}$, given that
\beq \frac{\delta \sin^2\theta_W}{\sin^2\theta_W}=\frac{\sin^2\theta_W{\cal W}+
\cos^2\theta_W{\cal Y}}{\sin^2\theta_W-\cos^2\theta_W}.\eeq
The measurement of $\sin^2\theta_{\rm W}$ with a total precision of $10^{-5}$
would determine ${\cal Y}$ with a
$\pm 3~10^{-5}$ precision, if ${\cal W}=0$.
If both ${\cal Y}$ and ${\cal W}$ are non-vanishing,
$\sin^2\theta_{\rm W}$ will restrict them to lie in a band, that becomes a long ellipse
taking into account the other measurements.


At LEP, the  LEP2 run above the $Z$ peak measured ${\cal W}$ and ${\cal Y}$ as well as the $Z$-peak LEP1 run~\cite{STWY},
because these  parameters give corrections that increase with the collider energy, e.g.\
\beq 
\frac{\sigma(e^+e^-\to \mu^+\mu^-)}{\sigma(e^+e^-\to \mu^+\mu^-)_{\rm SM}} = 1-  (0.67 {\cal W}+1.33 {\cal Y}) \frac{s}{M_W^2}\qquad\hbox{for $s\gg M_W^2$}.\eeq
Similarly,  we estimate that an $e^+e^-$ collider operating at higher energy $\sqrt{s}$
(around the $W^+ W^-, Zh$ and $t\bar t$ thresholds)
can measure ${\cal W,Y}$ with $\pm 0.3~10^{-4}$ accuracy~\cite{Janot}.

Furthermore, processes such as $e^+ e^- \to \gamma Z$ can probe the anomalous $\gamma\gamma Z$,
$\gamma ZZ$, etc vertices generated by a loop of heavy charged fermions or scalars.

\section{Conclusions}\label{concl}
We computed the maximal value of the width into $\gamma\gamma$ of a neutral scalar $S$ with mass $M_S$.

\smallskip

In section~\ref{psi} we considered the effect of a loop of charged fermions
with a Yukawa coupling $y$ to $S$.
Perturbativity of $y$ was quantified by computing the scale at which $y$ or any other
coupling, renormalised to higher energy, hits a Landau pole.
We also impose meta-stability bounds on the $S$ potential.
Fig.\fig{fermion} shows the maximal $\Ggg$ as function of the Landau pole scale.

\smallskip

In section~\ref{min} we  considered the effect of a loop of charged scalars
with a cubic coupling to $S$.
A large cubic does not lead to Landau poles, but it is, however, limited by
vacuum (meta)stability and perturbativity in a way that depends on dimensionless quartic couplings,
which are again subject to perturbativity bounds.
Meta-stability was computed considering the multi-field critical bounce.
Fig.\fig{scalargeneric} shows the maximal $\Ggg$ as function of the Landau pole scale.
The result is  similar to the fermionic case.

\medskip

In both the fermionic and the scalar case we allowed for $N$ states and considered the possibility that a new
$\SU(N)$ gauge symmetry acts on them.
The maximal value of $\Ggg$  allowed by perturbativity becomes qualitatively larger
if either the Yukawa coupling $y$ or the scalar cubic, in their renormalization group evolution,
can approach an infra-red fixed point.  
In such a case their maximal size is no longer controlled by Landau poles, but by the new gauge coupling $g$, 
which can be large.
Non-perturbative models discussed in the literature~\cite{750fits,nonpert} are recovered
in the limit where the new gauge coupling $g$ becomes non-perturbative around $M_S$.

\medskip

In section~\ref{col} we considered the connection with Dark Matter, finding
that $N\circa{>}2$ scalar doublets with mass $M\circa{<}M_S/2$
can thermally reproduce the cosmological DM abundance.
If they are quasi-degenerate, it becomes difficult to see them at hadronic colliders.
We discussed how  precision measurements can help in indirectly probing them.

\footnotesize

\subsubsection*{Acknowledgments}
This work was supported by the ERC grant NEO-NAT.
We thank Antonello Polosa for useful discussions.

\end{document}